\documentclass[10pt,conference,final,letterpaper]{IEEEtran}
\pdfoutput=1
\IEEEoverridecommandlockouts
\usepackage{cite}
\usepackage{nicefrac}
\usepackage{amsmath,amssymb,amsfonts}
\usepackage{algorithmic}
\usepackage{graphicx}
\usepackage{textcomp}
\usepackage{xcolor}
\usepackage{orcidlink}
\usepackage{tikz}
\usetikzlibrary{decorations.pathreplacing,calligraphy}

\usepackage{subcaption}
\def\BibTeX{{\rm B\kern-.05em{\sc i\kern-.025em b}\kern-.08em
    T\kern-.1667em\lower.7ex\hbox{E}\kern-.125emX}}

\begin{document}
\bstctlcite{IEEEexample:BSTcontrol}

\title{Towards Less Greedy Quantum Coalition Structure Generation in Induced Subgraph Games\\
\thanks{© 2024 IEEE. Personal use of this material is permitted.  Permission from IEEE must be obtained for all other uses, in any current or future media, including reprinting/republishing this material for advertising or promotional purposes, creating new collective works, for resale or redistribution to servers or lists, or reuse of any copyrighted component of this work in other works.\\The authors acknowledge funding from the German Federal Ministry of Education and Research, project Q-Grid, 13N16179 and 13N16177.}
}

\makeatletter
\newcommand{\linebreakand}{%
  \end{@IEEEauthorhalign}
  \hfill\mbox{}\par
  \mbox{}\hfill\begin{@IEEEauthorhalign}
}
\makeatother

\author{
\IEEEauthorblockN{Jonas Nüßlein\textsuperscript{$\orcidlink{0000-0001-7129-1237}$}}
\IEEEauthorblockA{\textit{LMU Munich}\\
jonas.nuesslein@ifi.lmu.de}
\and
\IEEEauthorblockN{Daniëlle Schuman\textsuperscript{$\orcidlink{0009-0000-0069-5517}$}}
\IEEEauthorblockA{\textit{LMU Munich}\\
danielle.schuman@ifi.lmu.de}
\and
\IEEEauthorblockN{David Bucher\textsuperscript{$\orcidlink{0009-0002-0764-9606}$}}
\IEEEauthorblockA{\textit{Aqarios GmbH}\\
david.bucher@aqarios.com}
\and
\IEEEauthorblockN{Naeimeh Mohseni\textsuperscript{$\orcidlink{0000-0003-3373-4572}$}}
\IEEEauthorblockA{\textit{\scalebox{.9}[1.0]{E.ON Digital Technology GmbH}}\\
naeimeh.mohseni@eon.com}
\linebreakand
\IEEEauthorblockN{Kumar Ghosh\textsuperscript{$\orcidlink{0000-0002-4628-6951}$}}
\IEEEauthorblockA{\textit{\scalebox{.9}[1.0]{E.ON Digital Technology GmbH}}\\
kumar.ghosh@eon.com}
\and
\IEEEauthorblockN{Corey O’Meara\textsuperscript{$\orcidlink{0000-0001-7056-7545}$}}
\IEEEauthorblockA{\textit{\scalebox{.9}[1.0]{E.ON Digital Technology GmbH}}\\
corey.o’meara@eon.com}
\and
\IEEEauthorblockN{Giorgio Cortiana\textsuperscript{$\orcidlink{0000-0001-8745-5021}$}}
\IEEEauthorblockA{\textit{\scalebox{.9}[1.0]{E.ON Digital Technology GmbH}}\\
giorgio.cortiana@eon.com}
\and
\IEEEauthorblockN{Claudia Linnhoff-Popien\textsuperscript{$\orcidlink{0000-0001-6284-9286}$}}
\IEEEauthorblockA{\textit{LMU Munich}\\
linnhoff@ifi.lmu.de}
}

\maketitle
\begin{abstract}
The transition to 100\% renewable energy requires new techniques for managing energy networks, such as dividing them into sensible subsets of prosumers called micro-grids. Doing so in an optimal manner is a difficult optimization problem, as it can be abstracted to the Coalition Structure Generation problem in Induced Subgraph Games, a NP-complete problem which requires dividing an undirected, complete, weighted graph into subgraphs in a way that maximizes the sum of their internal weights. Recently, Venkatesh et al.~\cite{venkatesh2023gcs} published a Quantum Annealing (QA)-based iterative algorithm called GCS-Q, which they claim to be the best currently existing solver for the problem in terms of runtime complexity. As this algorithm makes the application of QA to the problem seem promising, but is a greedy one, this work proposes several less greedy QA-based approaches and investigates whether any of them can outperform GCS-Q in terms of solution quality. While we find that this is not the case yet on D-Wave hardware, most of them do when using the classical QBSolv software as a solver. Especially an algorithm we call 4-split iterative R-QUBO shows potential here, finding all optima in our dataset while scaling favorably with the problem size in terms of runtime. Thus, it appears to be interesting for future research on quantum approaches to the problem, assuming QA hardware will become more noise-resilient over time.
\end{abstract}


\section{Introduction}

The adoption of an entirely renewable energy production, necessitated by the climate crisis, poses major challenges for energy network operators~\cite{lee2023ipcc, colucci2023power}: Weather-dependent, decentralized energy production by \textit{prosumers} makes load balancing in large energy grids more complicated~\cite{han2018incentivizing, colucci2023power}. One approach to solve this problem is partitioning these networks into self-sustained subunits, called \textit{micro-grids}~\cite{blenninger2024quantum, lasseter2004microgrid, han2018incentivizing}. But doing so in an optimal manner is a difficult optimization problem~\cite{lasseter2004microgrid, colucci2023power}: In their recent work, Mohseni et al.~\cite{mohseni2024competitive} show how this problem can be abstracted to the NP-complete \textit{Coalition Structure Generation problem (CSG)} in \textit{Induced Subgraph Games (ISGs)}~\cite{mohseni2024competitive, voice2012coalition}. Venkatesh et al.~\cite{venkatesh2023gcs} recently published a hybrid iterative algorithm for this problem using on \textit{Quantum Annealing (QA)}, a specialized type of quantum computing tailored to solving optimization problems. They claim it is currently the best algorithm to solve the problem in terms of runtime complexity~\cite{venkatesh2023gcs}. While efficient, the algorithm solves the problem greedily, possibly leading to suboptimal solutions. Thus, this paper collects several less greedy QA-based approaches to the problem – some adapted from literature, others contributed by us – and compares them to GCS-Q to see whether they outperform it in terms of solution quality. This is done by first explaining the problem formulation and related work, among which GCS-Q~\cite{venkatesh2023gcs}, in Sections~\ref{sec:background} \& \ref{sec:related}. Then, we detail our approaches in Section~\ref{sec:method} and evaluate them on a small dataset in Section~\ref{sec:experiments}, using both the classical software QBSolv as well as a D-Wave quantum annealer. Finally, we summarize these results and discuss their implications in Section~\ref{sec:conclusion}, suggesting to extend them in future work.
\section{Background}\label{sec:background}

Given a weighted, complete, undirected graph $\mathcal{G} = (N, E)$, containing a non-empty set $N$ of $n$ agents as nodes, the goal of solving the CSG in this ISG is to find the optimal \textit{coalition structure} $CS*$, which is defined as $CS* = \text{argmax}_{CS \in \mathcal{CS}(N)} V(CS)$ where $CS$ is a set of disjoint \textit{coalitions} $C$, $V(CS) = \sum_{C \in CS} \sum_{{i,j}\in C} w_{ij}$ is its value, $\mathcal{CS}(N)$ is the set of all possible coalition structures on $N$ and $w_{ij}$ is the weight of the edge $(i,j) \in E$ between the nodes, $i$ and $j$. Special types of coalitions are the \textit{grand coalition}, which contains all agents, and the \textit{singleton coalition}, which contains only one agent. This common version of the \textit{Clique Partitioning Problem (CPP)} is NP-complete, making it hard to solve with exact methods for larger sizes. Thus, we try to create improved approximate methods using QA.~\cite{bistaffa2021efficient, bistaffa2017algorithms, voice2012coalition, belyi2023subnetwork}

\section{Related Work}\label{sec:related}
The state-of-the-art in terms of QA-based algorithms designed to solve the CSG on ISGs is \textit{GCS-Q}~\cite{venkatesh2023gcs}. This hybrid approximation algorithm works by, starting from the grand coalition, iteratively splitting coalitions into two until this does not improve $V(CS)$ anymore. In each step of the algorithm, QA is used to solve a weighted minimum cut problem, formulated as a \textit{Quadratic Unconstrained Binary Optimization problem (QUBO)}, on the subgraph induced by the coalition $C$ to potentially split:

{\scriptsize
\begin{equation} \label{eq:qubo-gcsq-code}
    \sum_{i = 1}^{|C| - 1} \sum_{j = i + 1}^{|C|} \left( w_{ij} x_i + w_{ij} x_j + (- 2 w_{ij}) x_i x_j\right)
\end{equation}
}

Notice that always splitting a coalition into two is greedy and might lead to a suboptimal $CS$ in the end, e.g. if a 3-split would be optimal, but the agents needing to end up in a third coalition are assigned to different coalitions early on.
Determining experimentally that solving the QUBO with QA can be done in $\mathcal{O}(n)$, the authors conclude their algorithm's runtime is quadratic in $n$, and thus has a lower complexity than the leading classical approximation algorithm C-Link~\cite{farinelli2017hierarchical}, while exploring a larger proportion of the solution space. Since solving the splitting of the coalitions classically would take exponential time in $n$, the authors expect an advantage from running their algorithms on quantum hardware. When benchmarking the algorithm on D-Wave hardware against Tabu Search~\cite{palubeckis2004multistart, glover1986future}, Simulated Annealing~\cite{kirkpatrick1983optimization}, QBSolv~\cite{qbsolv} and Qiskit's NumpyMinimumEigensolver~\cite{numpyminimumeigensolver, mineigensolvertutorial, Qiskit} (which the authors refer to as “exact classical solver” or “Exact (ISG)”), Mohseni et al.~\cite{mohseni2024competitive} discern that GCS-Q using QA hardware finds solutions that are similar in quality to those found by all of the classical algorithms, while this QA-based algorithm scales better than all the classical ones in terms of runtime.~\cite{venkatesh2023gcs, venkatesh2023quacs, mohseni2024competitive}\\

Other relevant approaches to the problem include a QUBO-based algorithm by Kochenberger et al.~\cite{kochenberger2005clustering} (see Section~\ref{sec:method}), a similar, but hardware-specific QA-based approach by Leon et al.~\cite{leon2017multiagent, leon2019expressing} and the more generic quantum approaches BILP-Q~\cite{venkatesh2022bilp} and BOX-QUBO~\cite{nusslein2023black}. Relevant quantum approaches to related problems include ways to solve the weighted max-k-cut problem with QAOA~\cite{bako2022near, fuchs2021efficient, fakhimi2022max} and QA-based approaches for micro-grid creation, modeled as CSGs with additional constraints~\cite{colucci2023power} or a community detection problem~\cite{bucher2024evaluatingquantumoptimizationdynamic, blenninger2024quantum}.
\section{Methodology} \label{sec:method}

Our goal for this paper is to find or create QA-based approaches for CSG in ISGs that are less greedy than GCS-Q, and compare these approaches to this algorithm. To do this, we first compile a set of QUBO formulations of the problem, which we take or adapt from literature:

The first formulation is from work by Kochenberger et al.~\cite{kochenberger2005clustering} solving our problem using Tabu Search:

{\scriptsize
\begin{equation}\label{eq:kochenberger}
\left( \sum_{i = 1}^{n - 1} \sum_{j = i + 1}^{n} \left( - \, w_{ij} \sum_{c = 1}^{n} x_{ic} \, x_{jc} \right) \right) + \left(\sum_{i = 1}^{n} P \, (1 - \sum_{c = 1}^{n} x_{ic})^2 \right)
\end{equation}
}

This QUBO uses one-hot-encoded variables $x_{ic}$ to determine if an agent $i$ belongs to the coalition with the number $c$ or not. It rewards two agents being in the same coalition $c$ if they are connected by a high weight, while penalizing it if an agent belongs to several coalitions or none.
The QUBO can be solved as-is in one QA run to solve our given problem, thus constituting our first approach: \textit{Kochenberger}.

While the previous QUBO offers a suitable encoding of the problem, in case no domain knowledge about the maximum sensible number of coalitions is available, it uses the unnecessarily large amount $n$ of binary variables by requiring each agent $i$ to be part of exactly one coalition $c$: This amount is needed in the extreme case where the final $CS$ only contains singleton coalitions. The case where an agent is only part of a singleton coalition is, however, a fairly trivial one. Thus, we contribute another QUBO formulation that takes an agent $i$ to be part of its own singleton coalition if all its variables $x_{ic}$ have the value 0, and thus only needs $\lfloor \frac{n}{2} \rfloor$ variables – i.e. logical qubits on the quantum annealer – to encode the coalition numbers $c$ only for possible coalitions of at least size two. This halves the total number of logical qubits required. This formulation takes the form:

{\scriptsize
\begin{equation}\label{eq:zens}
\left( \sum_{i = 1}^{n - 1} \sum_{j = i + 1}^{n} \left( - \, w_{ij} \sum_{c = 1}^{\lfloor n/2 \rfloor} x_{ic} \, x_{jc} \right) \right)+ \left(\sum_{i = 1}^{n} P \, \sum_{c_1 = 1}^{{\lfloor n/2 \rfloor} - 1}\sum_{c_2 = c_1 + 1}^{\lfloor n/2 \rfloor} x_{ic_1}x_{ic_2} \right)
\end{equation}
}
Here, the first term again rewards respectively penalizes two agents being in the same coalition with the height of the weights connecting them, while the second term now only penalizes an agent $i$ being in more than one numbered coalition $c$, but allows it to be in less than one numbered coalition – namely as a zero-encoded singleton. We thus name this approach \textit{Zero-Encoded Singletons (ZEnS)}.

Another approach we introduce is \textit{n-split GCS-Q}. It adapts GCS-Q~\cite{venkatesh2023gcs} such that $CS*$ can also be found in one annealing run, by splitting the grand coalition into up to $n$ instead of up to two parts. This is done by combining the QUBO-formulation of the minimum cut problem from GCS-Q~\cite{venkatesh2023gcs} with the idea of using $n$ qubits per agent $i$ to one-hot-encode the numbered coalition the agent is part of from Kochenberger~\cite{kochenberger2005clustering}, creating the following weighted minimum $k$-cut formulation of our problem (with $k = n$), where each agent $i$ is to be part of exactly one coalition $c$:

{\scriptsize
\begin{equation}\label{eq:n-split-gcs-q}
    \sum_{i = 1}^{n - 1} \sum_{j = i + 1}^{n} \sum_{c = 1}^{n} \left( w_{ij} x_{ic} + w_{ij} x_{jc} + (- 2 w_{ij}) x_{ic} x_{jc}\right)+ \sum_{i = 1}^{n} P \, (1 - \sum_{c = 1}^{n} x_{ic})^2
\end{equation}
}

Furthermore, we also adapted a QUBO from literature on the weighted maximum $k$-cut problem~\cite{fakhimi2022max} to our problem by simply using the plain graph weights instead of their negation in the formula below – performing a minimum instead of a maximum cut – and again setting $k = n$:

{\scriptsize
\begin{align}\label{eq:r-qubo}
&\sum_{i = 1}^{n - 1} \sum_{j = i + 1}^{n} \sum_{c = 1}^{n-1} \left( w_{ij} x_{ic} + w_{ij} x_{jc} + (- 2 w_{ij}) x_{ic} x_{jc} - \sum_{c_2 \neq c} w_{ij} x_{ic} x_{jc_2}\right) \notag\\ &+ \left(\sum_{i = 1}^{n} P \, \sum_{c_1 = 1}^{n-2}\sum_{c_2 = c_1 + 1}^{n-1} x_{ic_1}x_{ic_2} \right)
\end{align}
}

This formulation, which we will call \textit{R-QUBO} (as the authors of~\cite{fakhimi2022max} do for their QUBO), follows a similar idea as $n$-split GCS-Q, but decreases the number of necessary variables per agent by one: Just like ZEnS, R-QUBO assigns a semantic to the case that all variables representing an agent $i$ have the value 0. But here, this bit-string is interpreted as the agent being part of a “coalition 0” that does not receive its own index number in the one-hot-encodings. To achieve this, the first part of the QUBO is adapted to ensure agents in “coalition 0” are treated the same as those in numbered coalitions, and the penalty constraint is modified to once more ensure that an agent is part of at most one, but not necessarily exactly one, numbered coalition.

The above-mentioned approaches all have in common that they consist of only a QUBO that captures our entire problem, and thus can find a complete solution by solving it with a single call to the quantum annealer. While this type of procedure makes the approaches completely non-greedy, it also requires quite large numbers of logical qubits, in $\mathcal{O}(n^2)$, which in the current era of noisy, intermediate-scale machines limits the scalability of the approaches in terms of numbers of agents $n$ for which the problem can still be embedded onto hardware and makes them more vulnerable to noise, as each additional qubit provides one more point where a bitflip-error can occur. As this means that the greediness and the number of required logical qubits of an approach form a trade-off, we decided also to create some algorithms that occupy an intermediate position between GCS-Q and the above-mentioned approaches, by making iterative versions of the latter that combine their QUBO-formulations with GCS-Q's way of iteratively splitting larger coalitions into smaller ones. This is done by choosing a number of possible sub-coalitions $k$ to split a given coalition into in one iteration, with $2 \leq k < n$ (or $2 \leq k < \lfloor n/2 \rfloor$ for ZEnS). (For iterative R-QUBO, this means splitting into $k-1$ numbered coalitions and one “0-coalition”.) This way, the resulting $k$-split algorithms are both less greedy compared to GCS-Q and less qubit-intensive compared to the non-greedy approaches.
\section{Experiments} \label{sec:experiments}

\subsection{Experimental setup}
We evaluate our approaches on an artificial toy dataset which was created based on simulations of real electricity grids. The dataset consists of graphs with even numbers of agents $n$, where $n \in [4, 28]$, and contains 20 graphs per $n$. 
We solve the CSG on ISGs with these graphs using both D-Wave's tabu-search-based classical meta-heuristic QBSolv~\cite{qbsolv, qbsolv_github}, as well as the D-Wave Advantage 4.1 quantum annealer~\cite{advantage}. For the iterative versions of our algorithms, we test $2 \leq k \leq \min(n, 27)$ with QBSolv and $2 \leq k \leq \min(n, 5)$ on the annealer. We restrict ourselves to the latter range on the D-Wave to save expensive Quantum Processing Unit (QPU), as this is the range we find the best results with QBSolv in and we already see clear downwards trend in terms of solution quality with increasing $k$ here (compare Fig.~\ref{fig:k}). 
In addition to the algorithms from Section~\ref{sec:method}, we also evaluate GCS-Q~\cite{venkatesh2023gcs}, to be able compare them to it. The source code for our experiments can be found at \url{https://github.com/Danielle-Schuman/Towards-Less-Greedy-Quantum-CSG-in-ISGs}.

To evaluate the quality of the solutions found by the algorithms, we compare their values to that of the optimal solution $CS*$, which we obtain using a classical exact solver created by Belyi et al.~\cite{belyi2023subnetwork}. We used two metrics:
One is the \textit{Number of Optima found (NO)} by a certain algorithm, i.e. if we are testing $a$ graphs of our dataset (the $a = 20$ graphs of the same size $n$), we look at the number $NO \leq a$ of solutions $CS_{\text{s}}$ for which $V(CS_{\text{s}}) = V(CS*)$.
The other metric is the \textit{Approximation Ratio (AR)},
which, for a particular solution $CS_{\text{s}}$, we define as
$AR = \frac{V(CS_{\text{s}})}{V(CS*)}$.
For both metrics, we averaged $V(CS_{\text{s}})$ over the 10 runs with the different random seeds when using QBSolv, and used 1 D-Wave run with 100 anneals per call otherwise.
We also measured the wall clock time taken by our algorithms. This can only be understood as an approximate measure of their runtime, given that we ran our experiments on machines we did not have exclusive access to. However, these approximate measurements still show interesting trends that justify future investigation of this aspect.

\subsection{Results}

\begin{figure*}[htbp]
  \begin{subfigure}{\columnwidth}
  \includegraphics[width=\textwidth]{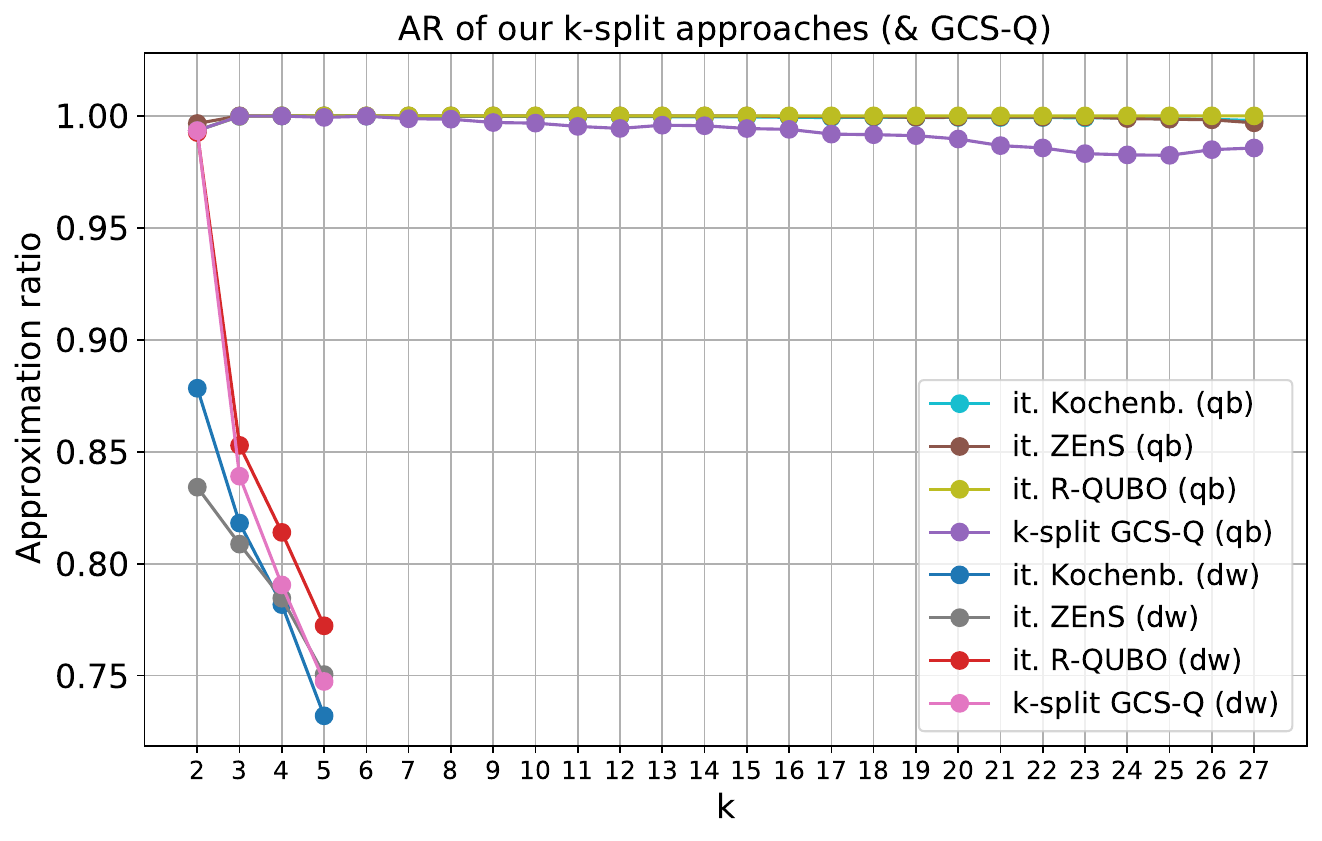}
  \caption{Average AR of our approaches \& GCS-Q.}
  \label{subfig:k-app}
  \end{subfigure}
  \hfill
  \begin{subfigure}{\columnwidth}
  \includegraphics[width=\textwidth]{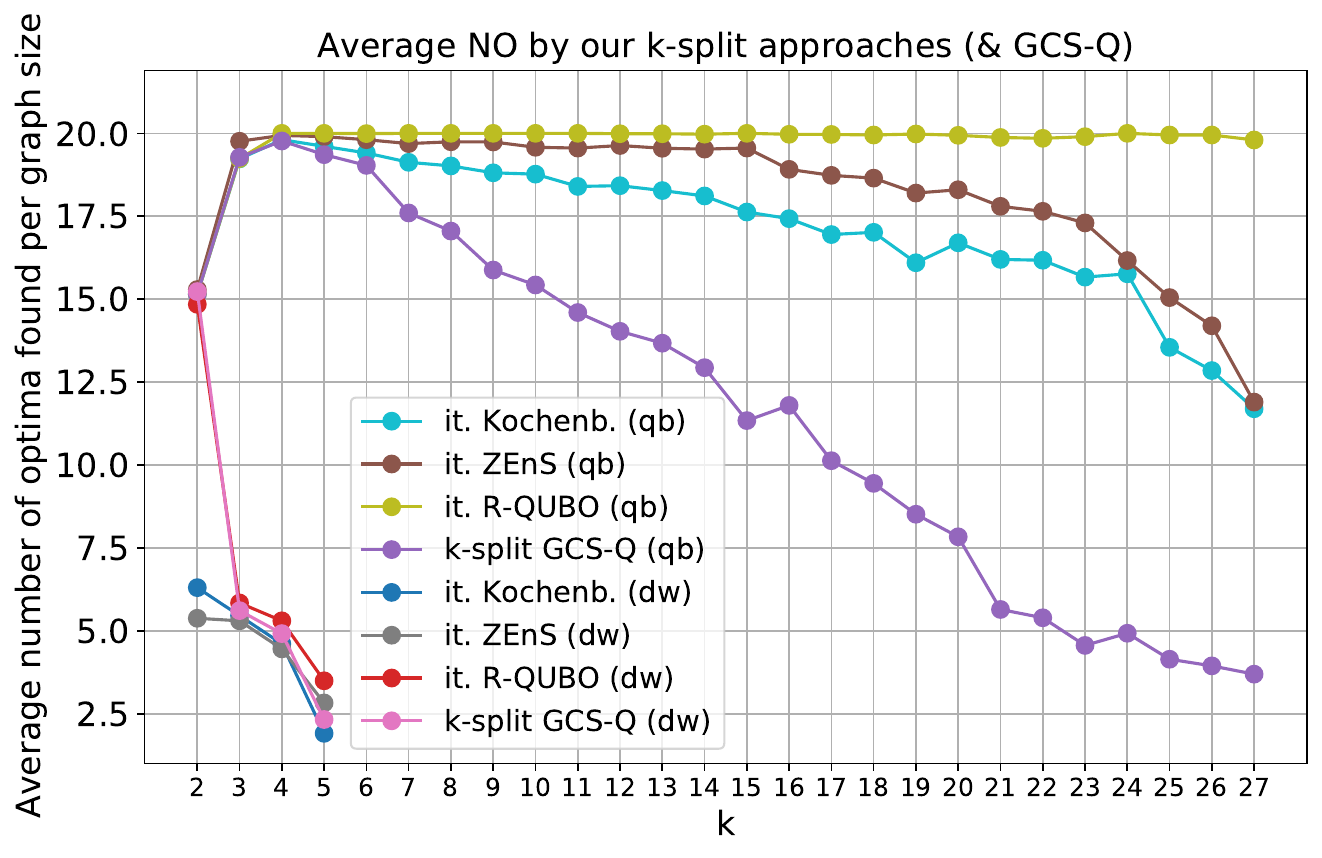}
  \caption{Average NO by our approaches \& GCS-Q.}
  \label{subfig:k-num}
  \end{subfigure}
  
  \caption{AR and NO averaged over all graphs in our dataset with $k \leq n$ (averaged first per $n$ and then over all $n \geq k$) for different values of $k$ for our algorithms and GCS-Q (shown as the value for $k=2$ of k-split GCS-Q). For QBSolv (qb), all values shown are averages over 10 differently seeded runs. For D-Wave (dw), only $k < 6$ were tried.}
  \label{fig:k} 
\end{figure*}

When evaluating our experiments, we found that with QBSolv, $k = 4$ is the best value of $k$ for all our iterative algorithms both in terms of AR as well as in terms of NO per graph size $n$, when averaging over our entire dataset (compare Fig.~\ref{fig:k})\footnote{We additionally compared the sums of all absolute values $V(CS_{\text{s}})$ we found given a certain $n$ for different values of $k$ to arrive at this conclusion. These numbers can be found in our repository.}. Figures~\ref{subfig:app-qbsolv} \& \ref{subfig:num-qbsolv} show the average ARs – averaged over the 20 graphs of each $n$ – as well as the NO per $n$, for both our 4-split iterative algorithms as well as our non-greedy algorithms and GCS-Q. As can be seen here, all algorithms perform very well with QBSolv, reaching an average AR of $> 0.9$ in all cases and finding a large NO at least for smaller $n$. Notice in particular how all algorithms except $n$-split GCS-Q outperform GCS-Q with regard to both metrics. We furthermore notice that the 4-split algorithms outperform their non-greedy counterparts for larger $n$, i.e. show improved scalability on our dataset. This becomes evident when comparing the NO in Fig.~\ref{subfig:num-qbsolv}. Here, 4-split iterative R-QUBO stands out especially, as it finds the optimum for all graphs in our dataset. As can be seen in Fig.~\ref{fig:time}, while GCS-Q still outperforms our best algorithms in terms of wall clock time taken, the 4-split algorithms do show a favorable scaling in regarding their runtime with respect to $n$.

\begin{figure*}[htbp]
  \begin{subfigure}{\columnwidth}
  \includegraphics[width=\textwidth]{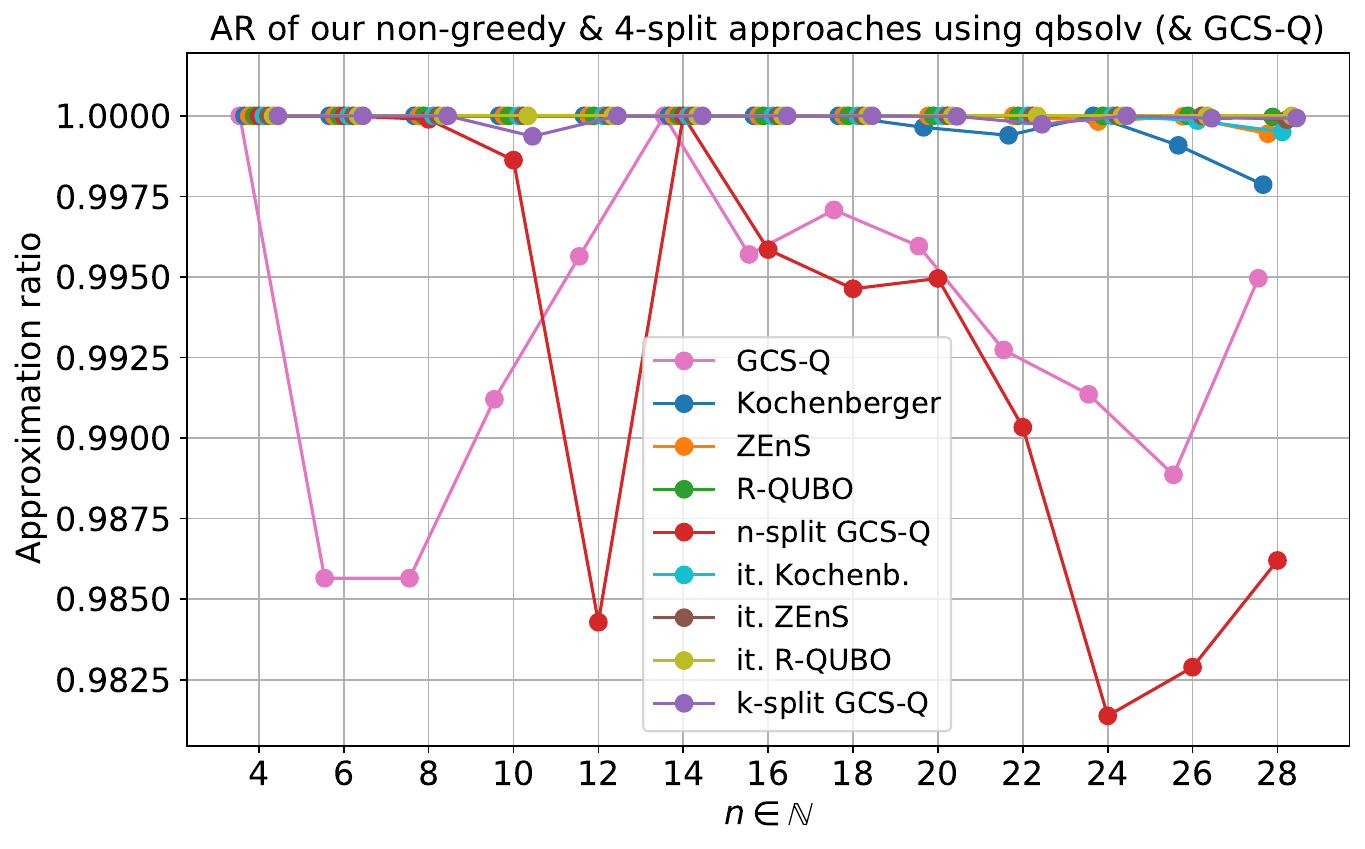}
  \caption{AR of our approaches \& GCS-Q using QBSolv.}
  \label{subfig:app-qbsolv}
  \end{subfigure}
  \hfill
  \begin{subfigure}{\columnwidth}
  \includegraphics[width=\textwidth]{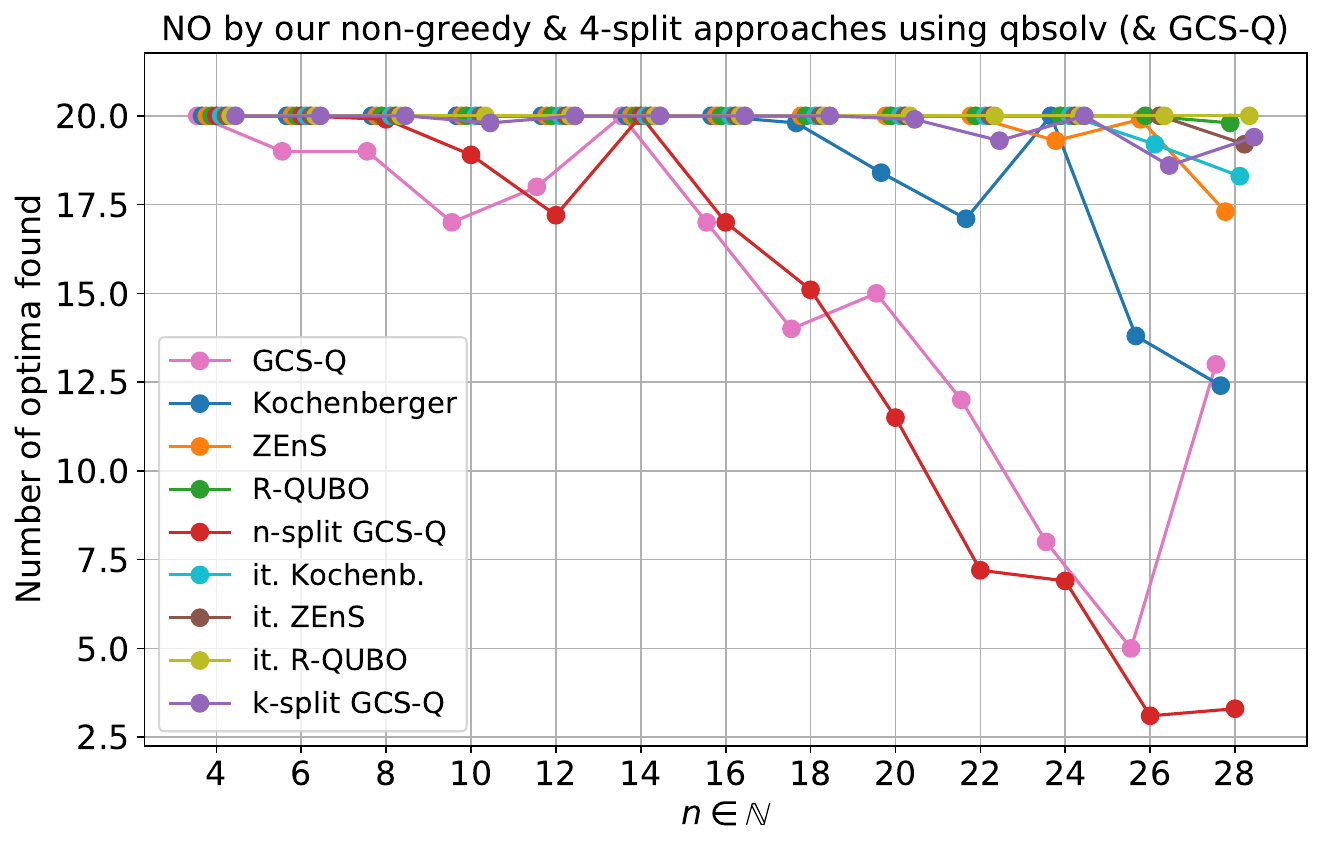}
  \caption{NO by our approaches \& GCS-Q using QBSolv.}
  \label{subfig:num-qbsolv}
  \end{subfigure}\\[0pt]
  
  \begin{subfigure}{\columnwidth}
  \includegraphics[width=\textwidth]{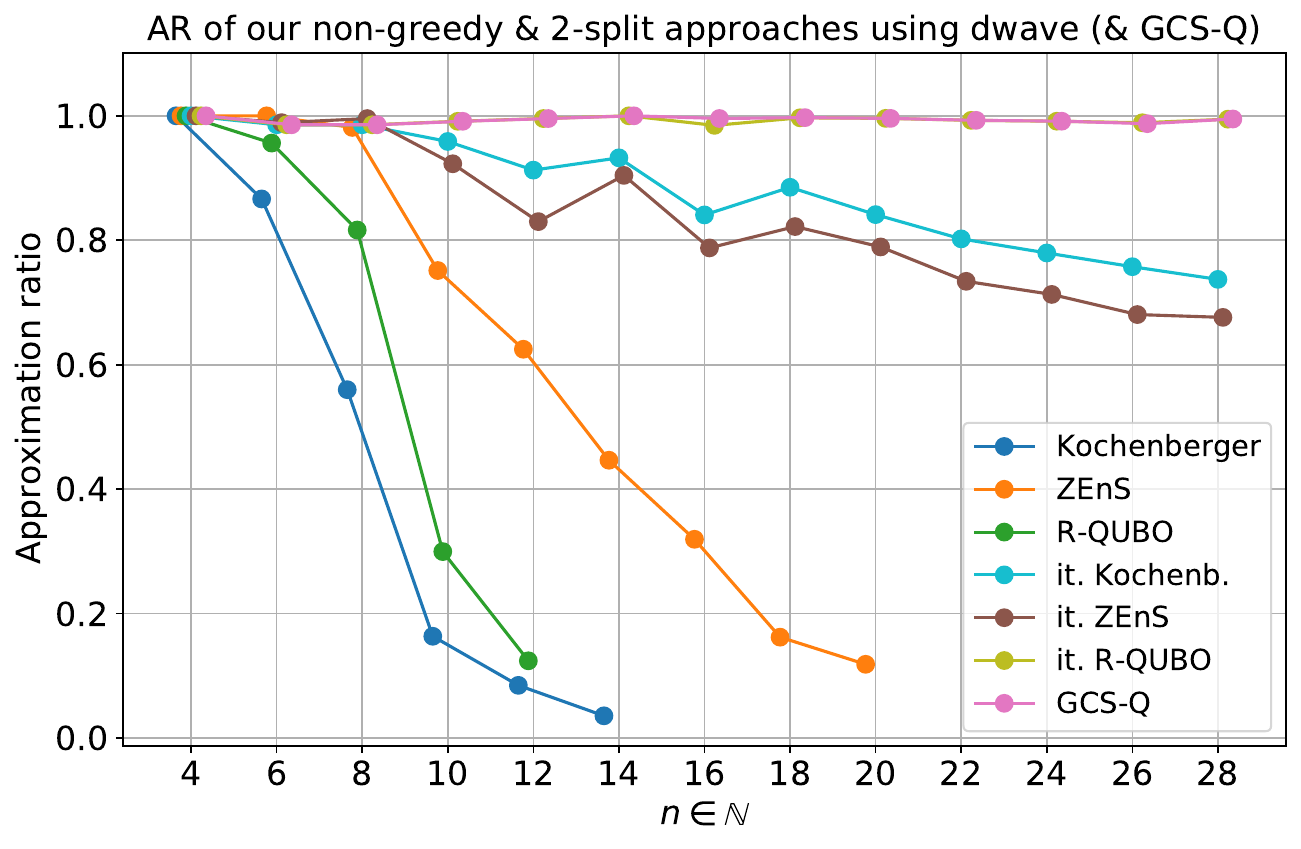}
  \caption{AR of our approaches \& GCS-Q using D-Wave.} 
  \label{subfig:app-dwave}
  \end{subfigure}
  \hfill
  \begin{subfigure}{\columnwidth}
  \includegraphics[width=\textwidth]{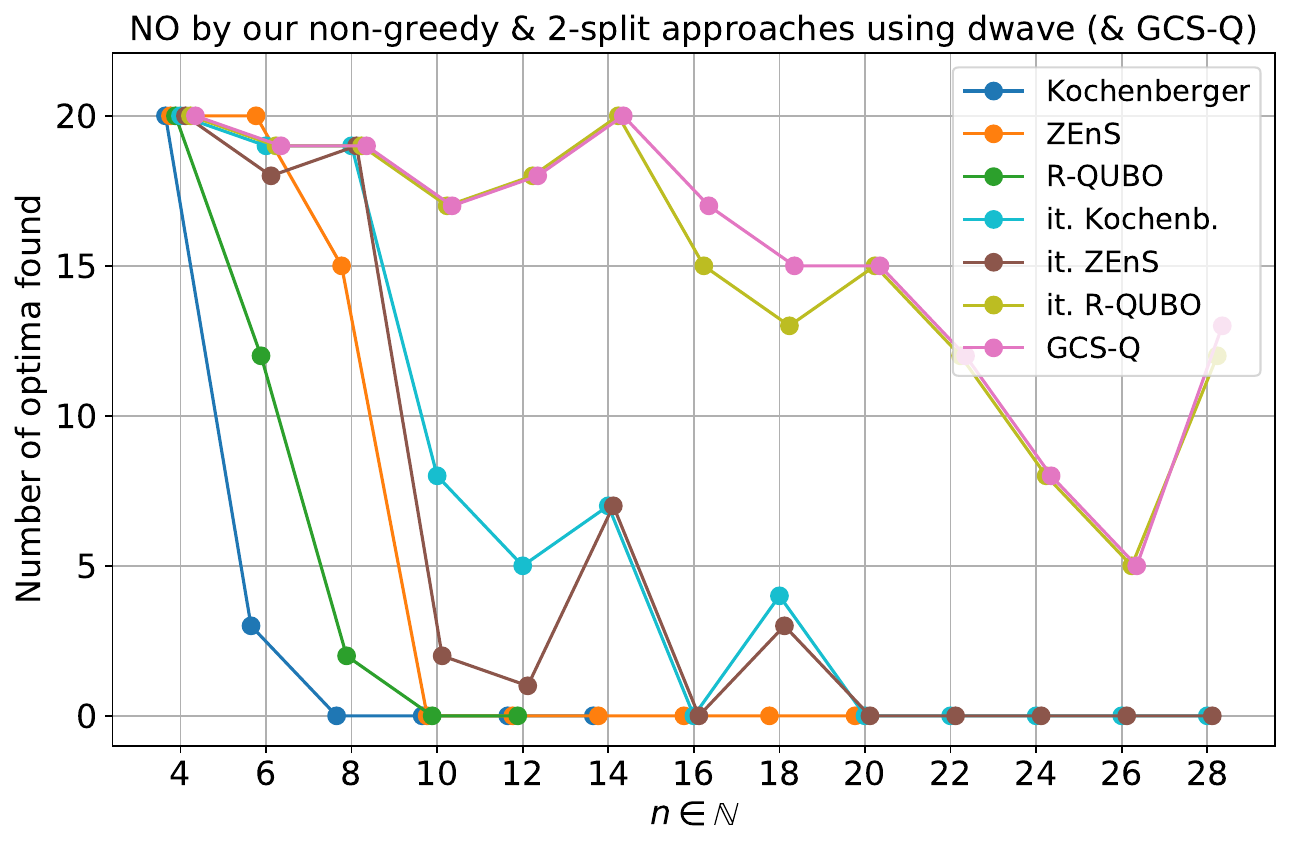}
  \caption{NO by our approaches \& GCS-Q using D-Wave.}
  \label{subfig:num-dwave}
  \end{subfigure}
  
  \caption{AR and NO per $n$ for our algorithms. For QBSolv, all values shown are averages over 10 differently seeded runs.}
  \label{fig:quality} 
\end{figure*}

When executing the algorithms using D-Wave hardware, we find very different results: Here, the best $k \in [2, 5]$ we find is $k = 2$ for all iterative algorithms (compare Fig.~\ref{fig:k}. In this case, iterative R-QUBO becomes the same algorithm as GCS-Q, due to the last and second-to-last terms in Eq.~\ref{eq:r-qubo} becoming empty, meaning any perceived differences between the two can only be attributed to noise or statistical fluctuations.
Furthermore – leaving out $n$-split GCS-Q because of its bad performance with QBSolv and $k$-split GCS-Q due to it following the same principle as GCS-Q 
– we find that GCS-Q outperforms all our (actually different) algorithms in terms of both our quality metrics for $n > 8$ when running with the D-Wave, compare Figures~\ref{subfig:app-dwave} \& \ref{subfig:num-dwave}. Additionally, we only find working embeddings onto the machine's hardware graph for all non-greedy approaches for $n \leq 12$ (compare e.g. Fig.~\ref{subfig:app-dwave}). Moreover, these algorithms occasionally find infeasible solutions (which were excluded when calculating the AR), i.e. ones breaking the constraint of every agent being assigned to at most one coalition, for $n \geq 8$. Regarding the wall clock time taken, the 2-split iterative approaches also perform best alongside GCS-Q. As can be seen in Fig.~\ref{fig:time}, the scaling with $n$ here seems less favorable in comparison to that with QBSolv, with the exception of 2-split ZEnS, which is the only algorithm consistently outperforming GCS-Q in this aspect.

\begin{figure}[htbp]
\captionsetup{width=\columnwidth}
\centerline{\includegraphics[width=\columnwidth]{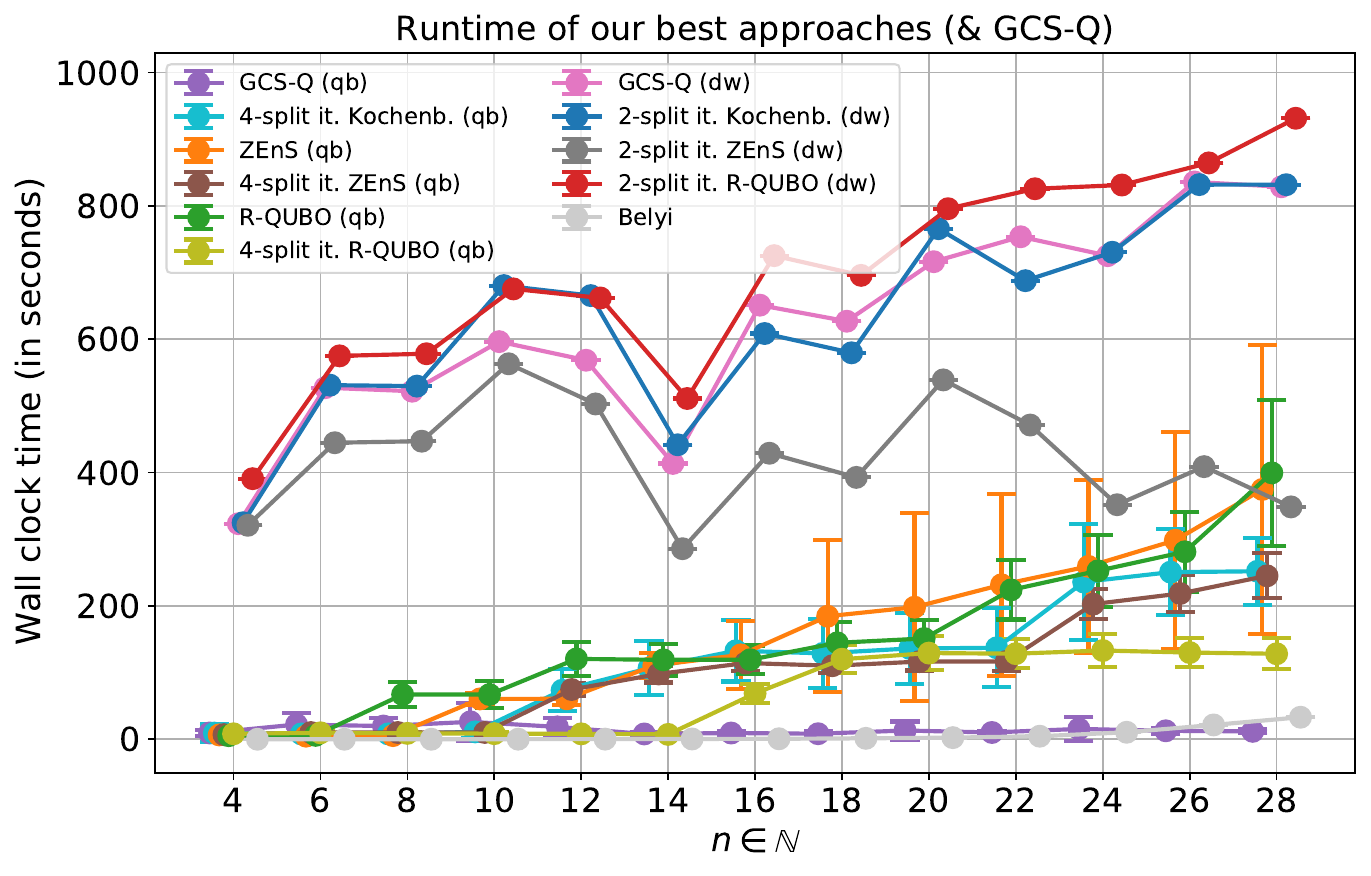}}
\caption{Wall clock time taken by our best algorithms and GCS-Q using QBSolv (qb) and D-Wave hardware (dw). The exact solver by Belyi et al.~\cite{belyi2023subnetwork} is shown for comparison of the scalability with $n$ (a direct comparison is not possible, as it was run on a different machine).}
\label{fig:time}
\end{figure}
\section{Discussion and Conclusion}\label{sec:conclusion}
In this work, we have investigated QA-based approaches to solve the CSG in ISGs, in order to find out if using less greedy approaches compared to GCS-Q can improve upon this QA-based state-of-the-art. We tested four different non-greedy, QUBO-based approaches – some taken from literature, some contributed by us – and four iterative approaches we designed by combining the non-greedy ones with the idea from GCS-Q to iteratively split given coalitions into more favorable ones, where we chose to split into a number of $k$ sub-coalitions per step, treating $k$ as a hyperparameter. When performing experiments on a small ($n \leq 28$) dataset, we found that none of our algorithms outperform GCS-Q in terms of solution quality when using D-Wave hardware, but almost all of them do when using the classical QBSolv solver. The algorithm \textit{4-split iterative R-QUBO} stands out, finding the optimum for all graphs in our dataset with QBSolv. We attribute these results to the notion that, while the iterative approaches introduce more greediness into the algorithm the smaller $k$ is, they also reduce the amount of logical qubits used in the QUBO formulation – and thus the search space for the solver and the opportunities to make bit-flip errors. The latter aspect is exacerbated on QA hardware due to the presence of noise and logical qubits being represented by several physical qubits: Fig.~\ref{fig:qubits} exemplifies the large differences in physical qubit count needed on the D-Wave by our different approaches, showing respective numbers for the non-greedy, 4-split and 2-split approaches that clearly decrease with $k$. Thus, our results with QBSolv point to the possibility that some of our algorithms could outperform GCS-Q with newer, more noise-resilient generations of hardware that provide increased qubit connectivity and thus need less physical qubits per logical one. Still, limiting the search space through iterative approaches might be beneficial. This is further corroborated by our preliminary time measurements, which indicate that the runtime of our iterative algorithms scales significantly better with respect to $n$ than that of our non-greedy approaches – especially on QA hardware, where the former is similar to GCS-Q's scaling. Iterative R-QUBO is the most promising of our algorithms, needing the smallest amount of logical qubits to perform a $k$-split.

\begin{figure}[htbp]
\captionsetup{width=\columnwidth}
\centerline{\includegraphics[width=\columnwidth]{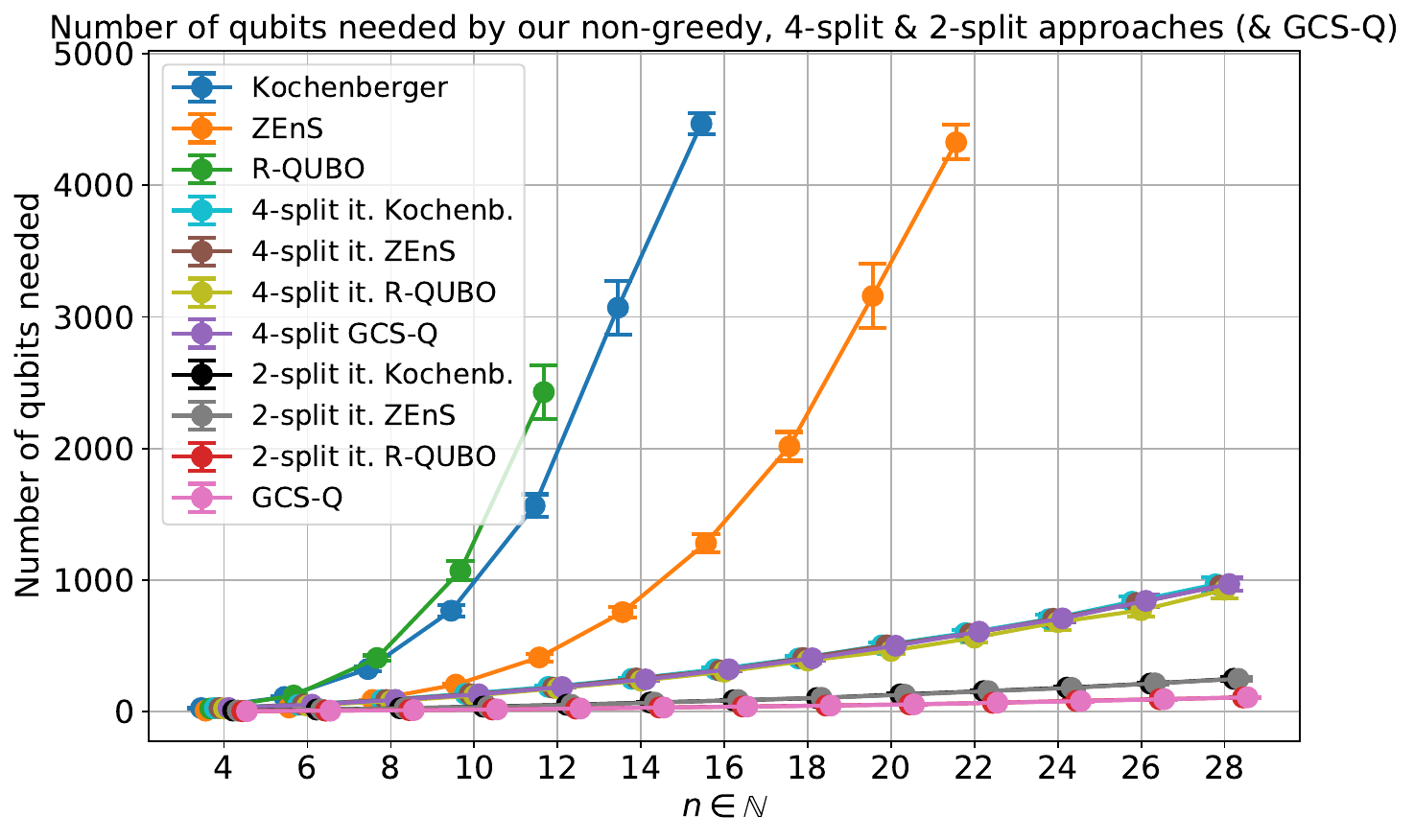}}
\caption{Number of physical qubits needed to run our non-greedy algorithms, our 4-split algorithms, our 2-split algorithms and GCS-Q on the D-Wave.}
\label{fig:qubits}
\end{figure}

While we expect this last-mentioned algorithm to be the most beneficial of our approaches on other datasets as well, given that it always “wins” the aforementioned trade-off by providing a certain level of greediness with the smallest amount of logical qubits, we cannot say with certainty how important a small amount of greediness would really be across datasets, and whether the optimal value for $k$ is dataset-specific – a circumstance we do not deem to be unlikely. Future work should thus include an evaluation of both the quantum-inspired versions of our iterative algorithms with QBSolv as well as the QA-based versions on different, larger datasets to investigate these circumstances, as well as scalability with larger $n$. Here, one should perform more D-Wave runs and more robust time measurements, to be able to compare the results to other approximate quantum~\cite{nusslein2023black} and especially classical algorithms. It would be interesting to see how our algorithms perform in comparison to other meta-heuristics like Simulated Annealing~\cite{keinanen2009simulated, hussin2017heuristic, gao2022improving, amorim1992clustering} or Tabu Search~\cite{hussin2017heuristic, sarkar2023p, kochenberger2005clustering, wang2006solving, wang2007clique, palubeckis2014iterated, amorim1992clustering}, and if they can compete with the classical state of the art. 
A next step would be to extend our set of quantum approaches by trying them with QAOA on a gate-based quantum computer, looking for more different QUBO approaches that use different types of encodings, e.g. unary encoding~\cite{Codognet22, Tamura21}, or introducing more real-world constraints from the domain of micro-grid creation~\cite{colucci2023power}. This way, we will come closer to determining the applicability of quantum approaches in real-world energy network management.

\bibliographystyle{IEEEtran} 
\bibliography{IEEEabrv,references}

\end{document}